# On the fluctuation-dissipation of the oxide trapped charge in a MOSFET operated down to deep cryogenic temperatures

G. Ghibaudo

IMEP-LAHC, Université Grenoble Alpes, MINATEC/INPG, 38016, Grenoble (France)
Email: gerard.ghibaudo@gmail.com

**Abstract**

We present an analysis of the oxide trapped charge noise in MOSFETs operated at deep cryogenic temperatures. Our study involves a revised derivation of the interface trap conductance Gp and the oxide trapped charge noise SQt at the SiO2/Si MOS interface under very low temperature conditions, considering the application of Fermi-Dirac statistics. We establish a new relationship between SQt and Gp, highlighting the limitations of the Nyquist relation at such low temperatures. Additionally, we introduce a novel formula for the oxide trapped charge 1/f noise that goes beyond the classical Boltzmann expression. This formula incorporates the oxide trap density and effective temperature, taking into account degenerate statistics.

**1. Introduction**

Low-frequency (LF) noise has been the subject of intensive research for many decades in MOS devices. It is now well-accepted that carrier number fluctuations (CNF) resulting from electron trapping/detrapping in the oxide are the main sources of LF noise in MOSFETs [1-6]. The LF noise in the channel thus originates from time variations in the oxide-trapped charge, which in turn induces fluctuations in the flat-band voltage and controls the inversion charge that regulates the drain current [5, 6]. The fluctuations in oxide-trapped charge are associated with energy dissipation due to electronic exchanges between the oxide traps and the band states, as described by the fluctuation-dissipation theorem [7, 8]. Indeed, the power spectral density of the oxide-trapped charge, SQt, was found to be directly proportional to the interface trap conductance, Gp, according to the Nyquist relation [7, 8]. It is worth noting that oxide-trapped charge noise has become a critical issue in quantum dot devices, where it could disturb the coherence time of silicon spin-qubits operation at very low temperatures [9, 10]. Moreover, oxide-trapped charge noise strongly limits the analog operation of CMOS circuits in the context of quantum computing, readout electronics, sensing, spatial electronics, etc. [11-15].

In this work, we propose a revised analysis of the oxide-trapped charge noise in a MOSFET operated at deep cryogenic temperatures. To this end, we carefully reconsider the derivation of the interface trap conductance, Gp, and oxide-trapped charge noise, SQt, at the SiO2/Si MOS interface under very low-temperature conditions, where Fermi-Dirac statistics should be used. We then reformulate the relationship between the oxide-trapped charge noise, SQt, and the interface trap conductance, Gp, demonstrating the inadequacy of the Nyquist relation [7, 8] at very low temperatures. Additionally, we establish a new formula for the 1/f noise of the oxide-trapped charge in terms of oxide trap density and effective temperature, taking into account degenerate statistics, thereby going beyond the classical Boltzmann expression.

**2. Dissipation and fluctuations of oxide trapped charge**

In this section, we first revisit the derivation of the interface trap conductance, Gp, which is associated with the dissipation resulting from electron exchange between the band and the traps. Next, we calculate the power spectral density of the trap occupancy function, which is associated with the fluctuations in the number of electrons in the traps, before establishing their relationship. We take care to perform the calculations using carrier Fermi-Dirac statistics to ensure applicability even under deep cryogenic conditions, where degenerate statistics may be relevant.

*2.1. Interface trap conductance*

The occupancy function, $f_t$, of a single-energy interface trap interacting with the conduction band follows the Shockley-Read-Hall statistics. The time variations of $f_t$ are described by the rate equation [16,17]:

$$\frac{df_t}{dt} = c_n \cdot n_s \cdot (1 - f_t) - c_n \cdot n_1 \cdot f_t, \qquad (1)$$

where $c_n$ is the capture rate, $n_s$ the carrier concentration and $n_1$ the concentration when the Fermi level $E_f$ equals the trap energy $E_t$. In a classical inversion layer, the capture rate is related to the trap cross section $\sigma$ and thermal velocity $v_{th}$ by $c_n = \sigma \cdot v_{th}$ [16], $n_s$ being the volume carrier concentration at the interface. In a quantized inversion layer, the capture rate for a single 2D subband reads $c_n = \sigma \cdot f_{esc}$, $f_{esc}$ being the escape frequency ($f_{esc} \approx 2 \cdot 10^{13}$ Hz) [18-20] and $n_s$ the areal carrier density.

The trap occupancy function in steady state i.e. when $df_t/dt=0$, $f_{t0}$, is therefore obtained from Eq. (1) as:

$$f_{t0} = \frac{n_{s0}}{n_{s0} + n_1}, \qquad (2)$$

where $n_{s0}$ refers to the carrier concentration at equilibrium. Note that, according to Eq. (2), $n_1$ corresponds to the carrier concentration when $f_{t0}=1/2$, i.e. when the trap is occupied at 50%. In terms of Fermi level, this means that $E_f$ crosses the trap energy, i.e. $E_f=E_t$.

Following Nicollian and Goetzberger [21] and within the small signal analysis, the variation of the trap occupancy function $\delta f_t = f_t - f_{t0}$ due to the AC signal applied on the gate $\delta V_g$, inducing a $\delta n_s$ ($=n_s-n_{s0}$) variation, is derived from Eq. (1), after having only kept the first order terms, in the form:

$$\frac{d\delta f_t}{dt} = -c_n \cdot (n_{s0} \cdot \delta f_t + n_1 \cdot \delta f_t + f_{t0} \cdot \delta n_s - \delta n_s) \,. \tag{3}$$

Taking the Fourier transform of Eq. (3) and solving for $\delta f_t$ yields as in [21],

$$\delta f_t = \frac{(1-f_{t0}) \cdot f_{t0}}{(1+j\omega\tau)} \cdot \frac{\delta n_s}{n_{s0}}. \tag{4}$$

where $\omega$ is the angular frequency, $j^2=-1$ and $\tau = f_{t0}/(c_n n_{s0})$.

The single trap admittance, $Y_s = i_s/\delta V_s$, with $i_s = j\omega \cdot q \cdot \delta f_t$ being the net current and $\delta V_s$ the surface potential variation, can thus be derived from Eq. (4) as in [21, Eq. (75)] in the form:

$$Y_s = j\omega \cdot q \cdot \frac{(1-f_{t0}) \cdot f_{t0}}{(1+j\omega\tau)} \cdot \frac{1}{n_{s0}} \cdot \frac{\delta n_s}{\delta V_s}. \tag{5}$$

If we further assume as in [21] that the Boltzmann statistics applies, we have $(\delta n_s/\delta V_s)/n_{s0}=q/kT$, $kT/q$ being the thermal voltage, so that taking the real part of Eq. (5) yields for the reduced single trap conductance $G_p/\omega$ the well-known expression [21, Eq. (17)]:

$$\frac{G_p}{\omega} = q \cdot \frac{(1-f_{t0}) \cdot f_{t0}}{kT/q} \cdot \frac{\omega\tau}{1+(\omega\tau)^2}. \tag{6}$$

However, since this expression is only valid under Boltzmann's statistics, which is not acceptable for low temperature operation, we should go one step further in the derivation with Eq. (5). To this end, we notice from Eq. (2) that we have $(1-f_{t0}) \cdot f_{t0} = n_{s0} \cdot (\delta f_{t0}/\delta n_{s0})$ and, after substitution in Eq. (5), we obtain for the reduced single trap conductance $G_p/\omega$,

$$\frac{G_p}{\omega} = q \cdot \frac{\omega\tau}{1+(\omega\tau)^2} \cdot \frac{\delta f_{t0}}{\delta V_s}. \tag{7}$$

At this stage, it is worth noticing that Eq. (7) can be derived in a more elegant way by rewriting Eq. (1) using Eq. (2) in the form,

$$\frac{df_t}{dt} = -\frac{f_t - f_{t0}}{\tau}. \tag{8}$$

Using the out-of-equilibrium trap occupancy function, $\delta f_t = f_t - f_{t0}$, the time differential equation can be written as,

$$\frac{d\delta f_t}{dt} + \frac{df_{t0}}{dt} = -\frac{\delta f_t}{\tau}. \tag{9}$$

Note that, in this formulation, $df_{t0}/dt$ refers to the quasi-stationary time variation of the trap occupancy function due to the surface potential variation with time $V_s(t)$, whereas $d\delta f_t/dt$ relates to its non-stationary counterpart. Then, proceeding within the small signal analysis as in Eq. (4) and taking the Fourier transform of Eq. (9) produces,

$$\delta f_t = -j\omega\tau \cdot \frac{1}{(1+j\omega\tau)} \cdot \frac{\delta f_{t0}}{\delta V_s} \cdot \delta V_s, \tag{10}$$

which provides for the single trap admittance $Y_s = i_s/\delta V_s$,

$$Y_s = q \cdot \frac{\omega^2 \tau}{(1+j\omega\tau)} \cdot \frac{\delta f_{t0}}{\delta V_s}. \tag{11}$$

Therefore, this derivation allows us to directly obtain the reduced single-trap conductance, $Gp/\omega$, as given by Equation (7). It is worth noting that in this approach, there is no requirement for a first-order expansion analysis of the rate equation, unlike in [21-23]. Furthermore, it highlights the clear distinction that can be made between the non-stationary and stationary contributions in the time variation of the trap occupancy function.

It is important to mention that, unlike Equation (6), the single trap admittance in Equation (11) and the reduced single trap conductance in Equation (7) are valid regardless of the statistics. These equations are embedded in the definition of ft0 in Equation (2) and should therefore be used in any low-temperature calculation where Boltzmann statistics are no longer applicable.

To evaluate the conductance of an energy continuum of states with a constant areal density, Nst (per eV per square centimeter), we follow the approach used in [21,24]. We integrate Equation (7) over trap energy, which is equivalent to integrating over surface potential or, in other words, integrating over ft0, which varies between 0 and 1. This integration results in the following expression:

$$\frac{G_p}{\omega} = q \cdot N_{st} \cdot \frac{\ln(1+\omega^2 \tau_{c0}^2)}{2\omega\tau_{c0}}, \tag{12}$$

where $\tau_{c0}=1/(c_n \cdot n_{s0})$ is the capture time. Interestingly, the conductance expression of Eq. (12) obtained for an energy continuum of states using Eq. (7) is the same as the one originally derived by Lehovec [24] and Nicollian and Goetzberger [21], when considering the Boltzmann statistics as in Eq. (6). Actually, this is because the kT/q factor appearing in Eq. (6) cancels out due to the integration process over energy.

Therefore, the conductance for an energy continuum of states, as given by Equation (12), is universal and applicable to both non-degenerate and degenerate carrier statistics. Hence, it can be used without modification for calculating the trap conductance of an energy continuum of states at both room and cryogenic temperatures. However, when dealing with single-energy traps, it is essential to use Equation (7) instead of Equation (6) for evaluating the trap conductance.

Moreover, for an energy continuum of states also uniformly distributed in the oxide and exchanging by tunneling with the conduction band, the overall reduced conductance can be obtained by integration over the oxide depth as [21, Eq. (104)]:

$$\frac{G_p}{\omega} = \int_0^{t_{ox}} q \cdot N_t \cdot \frac{\ln(1+\omega^2 \tau_{c0}^2 \cdot e^{2x/\lambda_{ox}})}{2\omega \tau_{c0} \cdot e^{x/\lambda_{ox}}} \cdot dx , \quad (13)$$

where $N_t$ is the volumetric oxide trap density (/eVcm$^3$) and $\lambda_{ox}$ ($\approx$0.1nm) the tunneling distance into the silicon oxide [25].

As can be seen in Fig. 1, Eq. (13) can be very well approximated by the analytical expression given by,

$$\frac{G_p}{\omega} = \frac{\pi}{2} \cdot q \cdot N_t \cdot \lambda_{ox} \cdot \frac{\omega \cdot \frac{\tau_{c0}}{\pi} \cdot e^{t_{ox}/\lambda_{ox}}}{\left(1+\omega \cdot \frac{\tau_{c0}}{\pi} \cdot e^{t_{ox}/\lambda_{ox}}\right)\left(1+\omega \cdot \frac{\tau_{c0}}{\pi}\right)} . \quad (14)$$

As a result, for an energy continuum of states uniformly distributed in the oxide, the reduced conductance $G_p/\omega$ is flat over a wide frequency range due to the huge tunneling time constant dispersion varying from $\tau_{c0}$ up to $\tau_{c0} \cdot \exp(t_{ox}/\lambda_{ox})$. We will see below that this plateau of $G_p/\omega$ vs frequency will be associated to a typical 1/f spectrum region for the oxide trapped charge fluctuations.

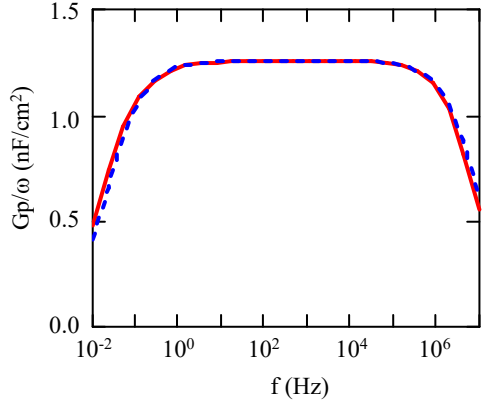

**Fig. 1.** Typical frequency variation of the reduced conductance Gp/ω for a continuum of states uniformly distributed into the oxide as obtained from the integral of Eq. (13) (red solid line) and the analytical approximation of Eq. (14) (blue dashed line) (parameters: $\tau_{c0}$=50ns, $\lambda_{ox}$=0.1nm, $t_{ox}$=2nm, $N_t$=5.10$^{17}$/eVcm$^3$).

2.1. Interface trap charge fluctuations

The power spectral density (PSD) of a single trap occupancy function, $f_t(t)$, can be obtained by applying the Wiener-Khintchine theorem to the auto-correlation function of $f_t(t)$, which is associated with the first-order kinetic process with a time constant, $\tau$, as described by Equation (8). This yields a Lorentzian spectrum for the PSD of $f_t(t)$, which is given by,

$$S_{ft} = 4 \cdot (\delta f_t)^2 \cdot \frac{\tau}{1+(\omega \cdot \tau)^2} , \quad (15)$$

where $(\delta f_t)^2$ is the full variance of $f_t(t)$ random variable. According to Machlup's analysis [26, see also 27], $f_t(t)$ is a two-state process varying randomly between 0 and 1 versus time, such that $(\delta f_t)^2 = f_{t0}(1-f_{t0})$. Therefore, the PSD of $f_t(t)$ can be written as,

$$S_{ft} = 4 \cdot f_{t0} \cdot (1 - f_{t0}) \cdot \frac{\tau}{1+(\omega \cdot \tau)^2} . \quad (16)$$

As was suggested in earlier works [7,8], it is meaningful to relate the PSD of the trap occupancy function $f_t(t)$ given by Eq. (16) to its associated conductance $G_p$ given by Eq. (7). Doing so, we obtain,

$$S_{ft} = 4 \cdot \frac{f_{t0} \cdot (1-f_{t0})}{\frac{\delta f_{t0}}{\delta V_s}} \cdot \frac{G_p}{q\omega^2} , \quad (17)$$

which, using Eq. (2), can be rewritten in the form,

$$S_{ft} = 4 \cdot \frac{1}{\frac{1}{n_{s0}} \cdot \frac{\delta n_{s0}}{\delta V_s}} \cdot \frac{G_p}{q\omega^2} . \quad (18)$$

When the Boltzmann statistics applies, we have $(\delta n_{s0}/\delta V_s)/n_{s0} = q/kT$, so that Eq. (18) becomes,

$$S_{ft} = 4 \cdot k \cdot T \cdot \frac{G_p}{q^2 \omega^2} . \quad (19)$$

Interestingly, this formulation corresponds to what was previously derived in [7,8] by considering the Nyquist relation for evaluating the power spectral density (PSD) of interface trap charge as a general consequence of the fluctuation-dissipation theorem. However, Equation (18) clearly indicates that the fluctuation-dissipation theorem or the Nyquist relation cannot be used in the most general case to calculate the PSD of interface trap charge, especially when Boltzmann statistics are not applicable. This is particularly relevant in cryogenic conditions and/or situations involving degenerate statistics.

Therefore, in the general case, the interface trap charge power spectral density $S_{Qt}$ must be calculated using Eq. (18), which can be reformulated under the elegant form,

$$S_{Qt} = 4 \cdot k \cdot T_{eff} \cdot \frac{G_p}{\omega^2} , \quad (20a)$$

with

$$T_{eff} = \frac{1}{k} \cdot \frac{\delta E_f}{\delta \ln(n_{s0})}, \quad (20b)$$

where $T_{eff}$ is an effective temperature and k is the Boltzmann constant. In the case of Boltzmann's statistics, $T_{eff}$ reduces to T, whereas, for e.g. a fully degenerate 2D subband, $T_{eff}$ depends on the Fermi energy as $T_{eff}=E_f/q$, $E_f$ being referenced to the subband edge.

For an energy continuum of states with constant areal density $N_{st}$ (/eVcm$^2$), using Eq. (12), we find for $S_{Qt}$ [in C$^2$/(Hz.cm$^2$)],

$$S_{Qt} = 4 \cdot k \cdot T_{eff} \cdot q \cdot N_{st} \cdot \frac{\ln(1+\omega^2 \tau_{c0}^2)}{2\omega^2 \tau_{c0}}. \quad (21)$$

Similarly, for an energy continuum of states uniformly distributed in the oxide and exchanging by tunneling with the conduction band, using the approximation of Eq. (14), we obtain for the interface trap charge power spectral density,

$$S_{Qt} = 4 \cdot k \cdot T_{eff} \cdot \frac{\pi}{2} \cdot q \cdot N_t \cdot \lambda_{ox} \cdot \frac{\frac{\tau_{c0}}{\pi} \cdot e^{t_{ox}/\lambda_{ox}}}{\left(1+\omega \cdot \frac{\tau_{c0}}{\pi} \cdot e^{t_{ox}/\lambda_{ox}}\right)\left(1+\omega \cdot \frac{\tau_{c0}}{\pi}\right)}. \quad (22)$$

## 3. Results and discussion

Figure 2 illustrates the typical variations of the reduced interface trap conductance, Gp/ω, with frequency, f, using Equations (7) and (12) for single-energy traps with various values of the occupation function, ft0, and for an energy continuum of traps, respectively. It is worth noting that the trap occupancy function, ft0, appearing in Equation (7), has been evaluated using Equation (2), considering a single 2D subband for the MOSFET inversion layer. As previously mentioned by Nicollian and Goetzberger [21, p.1093], the Gp/ω(f) characteristic for an energy continuum exhibits a slight deviation from a pure Lorentzian shape, showing some leveling at high frequencies compared to the curves for single-energy traps. Furthermore, it should be mentioned that the curve for the energy continuum of traps reaches a maximum value, Gp/ω|max ≈ 2.49·q·Nst, when the frequency is such that 2π·f·τ ≈ 1.98 [21, p.1093].

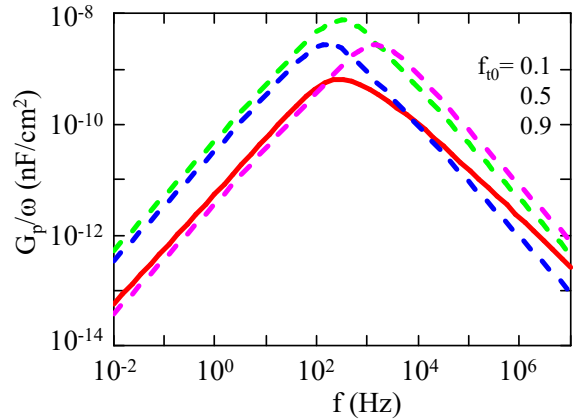

**Fig. 2.** Variations of reduced conductance $G_p/\omega$ with frequency f for single energy traps (dashed lines) with various occupancy function values $f_{t0}$ and for an energy continuum of traps (solid line) (parameters: $\tau_{c0}$=50ns, $\lambda_{ox}$=0.1nm, $t_{ox}$=2nm, $x_t$=0.5$t_{ox}$, T=300K, $N_{st}$=10$^{10}$/eVcm$^2$).

Figure 3 displays the typical frequency dependencies of the reduced conductance, Gp/ω, obtained using Equations (12) and (14) for an energy continuum of traps with various oxide depths, xt, and for an energy continuum of states uniformly distributed in the oxide. It can be observed that integrating Equation (12) over the oxide depth, from 0 to tox, results in the emergence of a plateau over a wide frequency range. The low and high cutoff frequencies of this plateau are equal to 1/[τc0·exp(tox/λox)] and 1/(2π·τc0), respectively. Additionally, it can be easily demonstrated from Equation (14) that the plateau reaches the value of π/2·q·Nt·λox. Importantly, this result aligns with both experimental and modeling findings by Uren et al. [28].

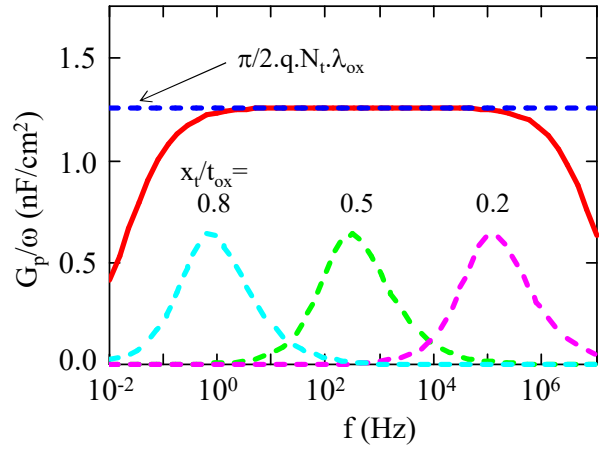

**Fig. 3.** Variations of reduced conductance $G_p/\omega$ with frequency f for an energy continuum of traps with various oxide depths $x_t$ (dashed lines) and for an energy continuum of states uniformly distributed in the oxide (solid line). The horizontal blue dashed line shows the analytical approximation value for the plateau region (parameters: $\tau_{c0}$=50ns, $\lambda_{ox}$=0.1nm, $t_{ox}$=2nm, $N_t$=5x10$^{17}$/eVcm$^2$).

Figure 4 depicts the frequency variations of the oxide trap charge power spectral density (PSD), SQt, for an energy continuum of traps with various oxide depths, xt, and for an energy continuum of traps uniformly distributed in the oxide, obtained using Equations (21) and (22) respectively. The 1/f noise spectrum, represented by q²kT.Nt.λox/f, corresponds to the plateau region of the Gp/ω(f) curve shown in Figure 3. This 1/f noise trend arises from the proportionality between SQt and Gp/ω^2 through Equation (20a). Figure 4 also illustrates how a 1/f noise spectrum can be constructed by the summation of multiple Lorentzian spectra with uniformly distributed cutoff frequencies in a logarithmic scale over a wide range. It is noteworthy that, in this example, the effective temperature, Teff, appearing in Equations (21) and (22), is close to the ambient temperature of 300K, as calculated using Equation (20b) with $n_{s0}$=10$^{11}$/cm$^2$.

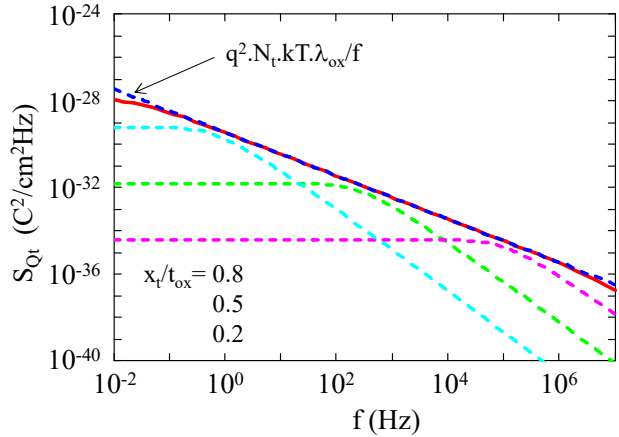

**Fig. 4.** Variations of the oxide trap charge PSD $S_{Qt}$ with frequency f for an energy continuum of traps with various oxide depths $x_t$ (dashed lines) and for an energy continuum of traps uniformly distributed in the oxide (solid line). The blue dashed straight line shows the 1/f trend corresponding to the $G_p/\omega$ plateau region (parameters: $\tau_{c0}$=50ns, $\lambda_{ox}$=0.1nm, $t_{ox}$=2nm, T=300K, $n_{s0}$=10$^{11}$/cm$^2$, $N_t$=5x10$^{17}$/eVcm$^2$).

Figure 5 demonstrates the typical temperature dependence of the oxide trap charge power spectral density (PSD), SQt, for an energy continuum of states uniformly distributed in the oxide, as obtained from Equation (22) at a fixed frequency of f = 100Hz. It is important to note that, in contrast to the 1/f noise spectrum given by the Boltzmann approximation (q².kT.Nt.λox/f, represented by the blue dashed line), SQt saturates at low temperatures, indicating the inadequacy of the Boltzmann limit, particularly below 30-40K. This deviation from the classical 1/f noise formula arises from the difference between the effective temperature, Teff, defined in Equation (20b) and used in Equation (22), and the ambient temperature, T, as demonstrated in Figure 6 for two different carrier areal densities.

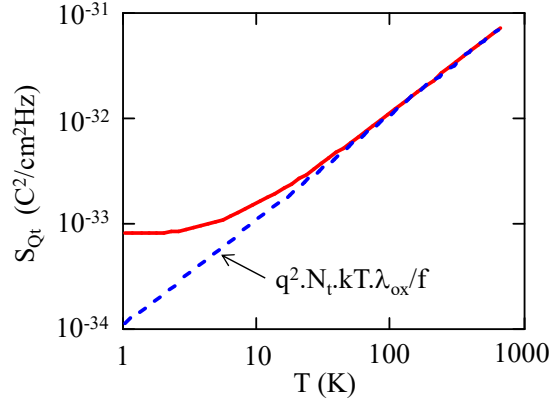

**Fig. 5.** Variations of the oxide trap charge PSD $S_{Qt}$ with temperature T for an energy continuum of states uniformly distributed in the oxide at a fixed frequency f=100Hz. The blue dashed line shows the 1/f noise trend given the Boltzmann approximation (parameters: $\tau_{c0}$=50ns, $\lambda_{ox}$=0.1nm, $t_{ox}$=2nm, $n_{s0}$=$10^{11}$/cm², $N_t$=5x$10^{17}$/eVcm²).

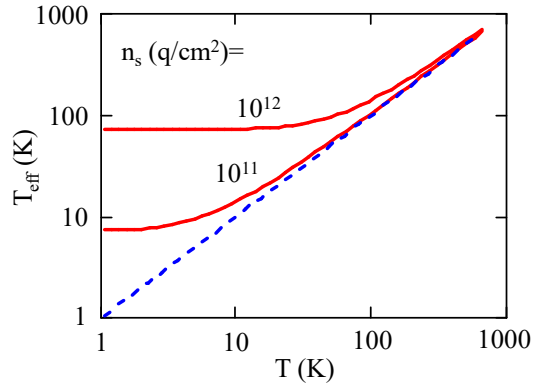

**Fig. 6.** Variations of the effective temperature $T_{eff}$ with temperature T obtained using Eq. (20b) for a 2D single subband with two carrier areal densities (parameter: 2D subband DOS $A_{2d}$=1.59x$10^{14}$/eVcm²).

As shown in Fig. 5, the oxide trap charge PSD $S_{Qt}$ strongly deviates at low temperature from the classical 1/f noise formula $q^2.kT.N_t.\lambda_{ox}/f$. Therefore, as it is usually done, it is meaningful to evaluate the trap density $N_{text}$ one could extract from the PSD $S_{Qt}$ by employing the classical Boltzmann 1/f formula such that,

$$N_{text} = \frac{S_{Qt} \cdot f}{q^2 \cdot \lambda_{ox} \cdot k \cdot T}. \quad \text{(Boltzmann statistics)} \qquad (23)$$

Figure 7 presents the variation of the normalized extracted trap density, $N_{text}/N_t$, with temperature and reciprocal temperature, as obtained using Equation (23) with SQt calculated using Equation (22), which accounts for the Fermi-Dirac statistics via Teff. It is evident that the trap density extracted using the classical trap charge 1/f noise formula diverges as 1/T below 60-80K. This signifies the limitation of the classical formula in accurately determining the trap density at lower temperatures.

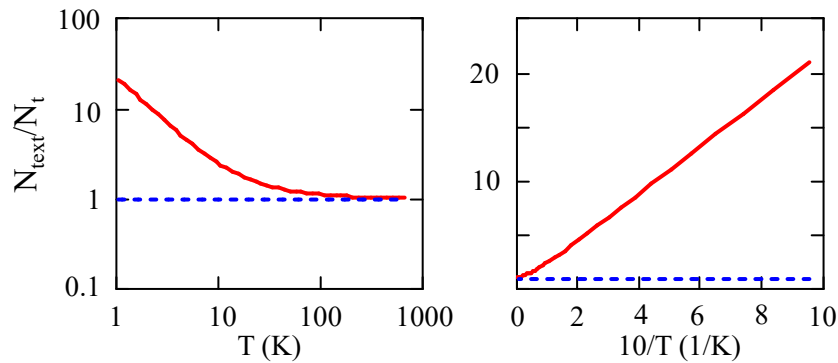

**Fig. 7.** Variation of the normalized extracted trap density $N_{text}/N_t$ with temperature T (left) and 10/T (right) as obtained with Eq. (23) with $S_{Qt}$ given by Eq. (22) for an energy continuum of states uniformly distributed in the oxide at a fixed frequency f=100Hz (parameters: $\tau_{c0}$=50ns, $\lambda_{ox}$=0.1nm, $t_{ox}$=2nm, $n_{s0}$=$10^{11}$/cm², $N_t$=5x$10^{17}$/eVcm²).

## 3. Summary and Conclusion

We have presented a revised analysis of the oxide trapped charge noise in MOSFETs operated at deep cryogenic temperatures. Our approach involved a careful reconsideration of the derivation of the interface trap conductance, Gp, and the oxide trapped charge noise, SQt, at the SiO2/Si interface, taking into account the utilization of Fermi-Dirac statistics at very low temperatures. We have also reformulated the relationship between SQt and Gp, highlighting the limitations of the Nyquist relation derived in previous works at such low temperatures. Furthermore, we have introduced a novel formula for the oxide trapped charge 1/f noise, which incorporates the oxide trap density and effective temperature, thereby extending beyond the conventional Boltzmann expression and accounting for degenerate statistics.